\documentclass{article}

\usepackage{hyperref}

\usepackage{amssymb}

\usepackage{amsmath}

\newcommand{\bea}{\begin{eqnarray}}

\newcommand{\eea}{\end{eqnarray}}

\newcommand{\be}{\begin{equation}}

\newcommand{\ee}{\end{equation}}

\newcommand{\rt}[1]{{}}

\newlength{\szovszel}
\newlength{\slashszel} 
\newcommand*{\sls}[1]{\mbox{%
    \settowidth{\szovszel}{\ensuremath{#1}}%
    \settowidth{\slashszel}{\ensuremath{\slash}}%
    \hspace*{0.5\szovszel}%
    \hspace*{-0.5\slashszel}%
    \slash%
    \hspace*{-0.5\szovszel}%
    \hspace*{-0.5\slashszel}%
    \ensuremath{#1}%
  }}

\begin{document}

\title{Local potential approximation\\ for the renormalization group flow of fermionic field theories}
\author{A. Jakov\'ac and A. Patk\'os\\
Institute of Physics, E\"otv\"os University\\
H-1117, P\'azm\'any P\'eter s\'et\'any 1/A, Budapest, Hungary}
\vfill
\maketitle
\begin{abstract}
The second functional derivative of the effective potential  of pure fermionic field theories is rewritten in a factorized form which facilitates the evaluation of the renormalization flow rate of the effective action in the Wetterich equation. It is applied to the local potential approximation in cases, when the effective potential depends on scalar composites built from the fermions. The procedure is demonstrated explicitly on the example of the $N_f$-flavor Gross-Neveu model and the one-flavor chiral Nambu--Jona-Lasinio model.   
\end{abstract}


\section{INTRODUCTION}

The functional renormalization group (FRG) \cite{wegner73,polchinski84,wetterich91,polonyi03} provides important insights for purely fermionic theories arising from integrating out the gluons in QCD \cite{wetterich96} or the Higgs sector of the electroweak Standard Model constructed with a composite Higgs field \cite{gies04}.
 Recent examples of the applications of FRG to pure fermionic field theories include the determination of the fixed point structure of the three-dimensional (3D) Thirring model \cite{gies10,janssen12} and of the 3D Gross-Neveu model \cite{scherer12a}, also for the case when the fermions interact with an external magnetic field \cite{scherer12b}. A thorough discussion of asymptotic safety has been presented for this model \cite{braun11}. 

The trial effective action commonly chosen for these investigations contains only the lowest powers of the invariant expressions formed from the fermi fields. The RG flow is then determined by solving a coupled set of first order evolution equations for the coupling strengths appearing in front of these lowest dimensional operators. In order to go beyond these Ans\"atze eventually bosonic auxiliary fields are introduced in the partial bosonization (Hubbard-Stratonovich transformation). The auxiliary fields are constructed to form irreducible multiplets under the space-time and internal symmetries of the model. The dependence of the effective potential on certain invariant combinations of the auxiliary fields, which can serve as order parameters, can be determined with the help of the FRG equations, in principle without restricting its functional form. Since the bosonization is only partial, one is faced with the dilemma of whether some redundant direct dependence of the effective action remains on the combinations of the fermi fields which share quantum numbers  with the bosonic invariants (e.g. dependence on higher powers of the composite fermionic operators). A similar redundancy was discussed in \cite{gies02} concerning the four-fermion coupling in the partially bosonized Nambu--Jona-Lasinio (NJL) model. Besides the direct coupling, the exchange of the auxiliary boson also contributes to the strength of this operator. The solution chosen in \cite{gies02} was to continuously tune the Hubbard-Stratonovich transformation along the renormalization flow in a way that forces the direct four-fermion coupling to vanish at all scales. 

The purpose of the present investigation is to avoid in a most straightforward manner the ambiguity originating from a simultaneous representation of the composite operators by the corresponding auxiliary bosonic and the original fermi fields. In the spirit of the Hubbard-Stratonovich identity we propose to work out the FRG equation in a space-independent Grassmannian background $(\zeta^\alpha, \bar\zeta^\alpha)$ characterized by nonzero values for a set of bilinear fermion composites $\bar\zeta^\alpha\Gamma^{\alpha\beta}_n\zeta^\beta$ fully contracted in the spinor indices $\alpha$ (with $\Gamma^{\alpha\beta}_n$ denoting generalized Dirac matrices). This assumption will allow us to introduce an effective potential depending on arbitrary high powers of the set of those invariants that are compatible with the space-time and internal symmetries of a given model. Our approach reflects an alternative strategy to purely fermionic theories compared to earlier investigations. The authors of Ref.\cite{comellas97} did not consider any condensation alternative
for scalar composites built from the fermions and truncated the Ansatz of the effective action at the highest power allowed by the Clifford algebra of the Grassmann fields. This is a valid approach above the compositeness scale \cite{gies02}, or for $d= 2$ dimensional theories with continuous internal symmetry in general. Then one can search for nontrivial fixed points involving operators with higher
(second and third) derivatives of the fermi fields which also imply nontrivial wave function renormalization ($\eta\neq 0$).

We shall construct the renormalisation group equations (RGE) in the local potential approximation (LPA) for the effective potential depending on a set of specific composites of bosonic nature built from the fermi fields without introducing any auxiliary variable. The central element of this construction will be a factorization of the $\Gamma^{(2)}$ matrix of the second functional derivatives of the effective quantum action $\Gamma$ with the help of an appropriately designed propagator. This step makes  the evaluation of the functional determinant of the Grassmann-valued functional matrix more efficient. One still has to keep in mind that the final result for the rate of variation of the effective potential in the Wetterich equation might depend on a larger set of field combinations formed from the constant spinorial background, and invariant under Lorentz and internal transformations compared to the set upon which the effective potential is assumed to depend in the LPA Ansatz. If this happens one arrives at an unambiguous set of equations \cite{jaeckel03} by eventually selecting a set of independent combinations, also taking into account Fierz's rearrangements. Then the flow can be restricted in a self-consistent way to the subset of operators that were taken into account from the start in the construction of $\Gamma^{(2)}$. 

Our discussion will proceed along the following lines. In Sec. 2 the general analysis is described without specifying the fermionic model in question. The central result of this paper is a fully general factorization formula for $\Gamma^{(2)}$. The first detailed application is presented in Sec. 3, where a LPA RGE is explicitly derived for the Gross-Neveu model. In Sec. 4 the essentials of the derivation are outlined for the chiral one-flavor Nambu--Jona-Lasinio model. Section 5 contains a short summary of results with an outlook for further applications. More involved details of the matrix algebra are presented in two appendixes.

\section{FACTORIZATION OF THE FERMIONIC FUNCTIONAL DETERMINANT IN THE RGE} 
 
The Wetterich equation describing the evolution of the effective action with the momentum scale $k$ is of the form
\be
\partial_k\Gamma_k=\frac{1}{2}{\textrm {Str}}\left[\partial_kR_k(\Gamma^{(2)}_k+R_k)^{-1}\right]=\frac{1}{2}\hat\partial_k{\textrm {Str}}\log(\Gamma^{(2)}_k+R_k),
\label{wetterich-equation}
\ee
where $\Gamma_k$ denotes the effective action resulting from integrating over all quantum fluctuations above the momentum scale $k$ using a convenient suppression function $R_k$ for the modes below the scale $k$. The derivative $\hat\partial_k$ denotes an operation consisting of first taking  the derivative of the logarithm with respect to $R_k$ and multiplying the result by $\partial_kR_k$.
The super-trace operation $\textrm{Str}$ is performed on functional matrices defined over the space of the Fourier-transformed extended column vectors of the fermi fields $\Psi(q)=(\psi(q), \bar\psi^T(-q))$ (discrete flavor indices are understood implicitly) and the corresponding transposed row vectors $\Psi^T(q)=(\psi^T(-q),\bar\psi(q))$. This doubling of the fermionic degrees of freedom follows the calculational setup of Ref.\cite{gies02}. The functional second derivative matrix is defined as
\be
\Gamma^{(2)}=\frac{\overrightarrow{\delta}}{\delta\Psi^T}\Gamma\frac{\overleftarrow{\delta}}{\delta\Psi}
\ee
and below we denote its elements as
\be
\Gamma^{(2)}_{\psi^T\psi}=\frac{\overrightarrow{\delta}}{\delta\psi^T}\Gamma\frac{\overleftarrow{\delta}}{\delta\psi},\quad
\Gamma^{(2)}_{\bar\psi\psi}=\frac{\overrightarrow{\delta}}{\delta\bar\psi}\Gamma\frac{\overleftarrow{\delta}}{\delta\psi},\quad
\Gamma^{(2)}_{\psi^T\bar\psi^T}=\frac{\overrightarrow{\delta}}{\delta\psi^T}\Gamma\frac{\overleftarrow{\delta}}{\delta\bar\psi^T},\quad
\Gamma^{(2)}_{\bar\psi\bar\psi^T}=\frac{\overrightarrow{\delta}}{\delta\bar\psi}\Gamma\frac{\overleftarrow{\delta}}{\delta\bar\psi^T}.
\ee

If there were no nonzero entries $\Gamma^{(2)}_{\psi^T\psi}$ and $\Gamma^{(2)}_{\bar\psi\bar\psi^T}$ one could calculate the  "tracelog" easily for the two independent Dirac-like kernels. Adding the two equal contributions (each divided by 2) and providing the extra fermionic minus sign one arrives at a usual fermionic one-loop formula. This observation suggests  that we construct a factorized form for the hypermatrix:
\be
\Gamma^{(2)}_k+R_k=(I+U_L)(\tilde\Gamma^{(2)}_k+R_k)(I+U_R),
\label{factorisation}
\ee
having the features
\be
\tilde\Gamma^{(2)}_{\psi^T\psi}=\tilde\Gamma^{(2)}_{\bar\psi\bar\psi^T}=0. 
\ee
The tracelog is performed for the three factors additively, and one finds that the evaluation becomes quite transparent  for the factors containing $U_L$ and $U_R$.

For the determination of the factorized form (\ref{factorisation}) consider the product of the following three matrices:
\be
\begin{pmatrix}
{1} & {\frac{1}{2}\Gamma^{(2)}_{\psi^T\psi}\tilde G_\psi}\\
{\frac{1}{2}\Gamma^{(2)}_{\bar\psi\bar\psi^T}\tilde G_\psi^{(T)}}&
{1}
\end{pmatrix}
\cdot
\begin{pmatrix}
{0} & {\tilde G_\psi}^{(T)-1}\\
{\tilde G_\psi^{-1}}&
{0}
\end{pmatrix}
\cdot
\begin{pmatrix}
{1} & {\tilde G_\psi}\frac{1}{2}\Gamma^{(2)}_{\bar\psi\bar\psi^T}\\
{\tilde G_\psi^{(T)}\frac{1}{2}\Gamma^{(2)}_{\psi^T\psi}}&
{1}
\end{pmatrix}.
\ee
We require the product to be equal to the original matrix.
This is fulfilled for the diagonal elements independently of the choice of $\tilde G_\psi$ and $\tilde G_\psi^{(T)}$. The equality of the off-diagonal blocks leads to requirements that determine these two effective propagators:
\bea
\Gamma^{(2)}_{\psi^T\bar\psi^T}&=&\tilde G_\psi^{(T)-1}+\frac{1}{4}\Gamma^{(2)}_{\psi^T\psi}\tilde G_\psi\Gamma^{(2)}_{\bar\psi\bar\psi^T},\nonumber\\
\Gamma^{(2)}_{\bar\psi\psi}&=&\tilde G_\psi^{-1}+\frac{1}{4}\Gamma^{(2)}_{\bar\psi\bar\psi^T}\tilde G_\psi^{(T)}\Gamma^{(2)}_{\psi^T\psi}.
\label{consistency}
\eea

Once the new fermion propagators are determined one can calculate the tracelog for all three factors. In particular (exploiting the permutation invariance of the trace operation) one finds
\be
\displaystyle
-\frac{1}{2}\left({\textrm {Tr}}\log(I+U_L)+{\textrm {Tr}}\log(I+U_R)\right)
=-{\textrm {Tr}}\log\left(I-\frac{1}{4}\Gamma^{(2)}_{\psi^T\psi}\tilde G_\psi\Gamma^{(2)}_{\bar\psi\bar\psi^T}\tilde G_\psi^{(T)}\right)
\ee
and the general form of the Wetterich equation for pure fermionic systems after this factorization reads as
\be
\partial_k\Gamma_k=-\frac{1}{2}\hat\partial_k{\textrm {Tr}}\left[\log(\tilde G^{-1}_{R,k})+\log(\tilde G^{(T)-1}_{R,k})+2\log\left(I-\frac{1}{4}\Gamma^{(2)}_{\psi^T\psi}\tilde G_{R,k}\Gamma^{(2)}_{\bar\psi\bar\psi^T}\tilde G_{R,k}^{(T)}\right)\right],
\label{wetterich-factorized}
\ee
where $\tilde G_{R,k}$ denotes the effective fermion propagator supplemented with the infrared cutoff function $R_k$.

Our central result will be an expression of the right-hand side of
(\ref{wetterich-factorized}) written explicitly in terms of the
spinor-valued matrix elements of $\Gamma^{(2)}$. First of all one can
solve (\ref{consistency}) for the effective propagators (for details
see Appendix A):
\bea
\label{eq:contr_solve}
&\displaystyle
\tilde G_\psi=2\Gamma^{(2)-1}_{\psi^T\psi}\left(1-\sqrt{1-N}\right)\Gamma^{(2)}_{\psi^T\bar\psi^T}\Gamma^{(2)-1}_{\bar\psi\bar\psi^T},\nonumber\\
&\displaystyle
\tilde G^{(T)}_\psi=2\Gamma^{(2)-1}_{\bar\psi\bar\psi^T}\Gamma^{(2)}_{\bar\psi\psi}\Gamma^{(2)-1}_{\psi^T\psi}\left(1-\sqrt{1-N}\right),
\eea
where we use the notation
\be
N=\Gamma^{(2)}_{\psi^T\psi}\Gamma^{(2)-1}_{\bar\psi\psi}\Gamma^{(2)}_{\bar\psi\bar\psi^T}\Gamma^{(2)-1}_{\psi^T\bar\psi^T}.
\ee
Next, one can express with the help of these propagators the terms on the right-hand side of (\ref{wetterich-factorized}) as follows:
\bea
2{\textrm {Tr}}\log\left(I-\frac{1}{4}\Gamma^{(2)}_{\psi^T\psi}\tilde G_{R,k}\Gamma^{(2)}_{\bar\psi\bar\psi^T}\tilde G_{R,k}^{(T)}\right)&=&2{\textrm {Tr}}\log(2N^{-1})+{\textrm {Tr}}\log(1-N)+2{\textrm {Tr}}\log\left(1-\sqrt{1-N}\right),\nonumber\\
{\textrm {Tr}}\log(\tilde G^{-1}_{R,k})&=&-{\textrm {Tr}}
\log\left(2\Gamma^{(2)-1}_{\psi^T\psi}\Gamma^{(2)}_{\psi^T\bar\psi^T}\Gamma^{(2)-1}_{\bar\psi\bar\psi^T}\right)-{\textrm {Tr}}\log\left(1-\sqrt{1-N}\right),\nonumber\\
{\textrm {Tr}}\log(\tilde G^{(T)-1}_{R,k})&=&-{\textrm {Tr}}\log\left(2\Gamma^{(2)-1}_{\bar\psi\bar\psi^T}\Gamma^{(2)}_{\bar\psi\psi}\Gamma^{(2)-1}_{\psi^T\psi}\right)-{\textrm {Tr}}\log\left(1-\sqrt{1-N}\right).
\eea
It is remarkable that the terms containing a square root mutually cancel in (\ref{wetterich-factorized}). After further cancellations one arrives at the rather simple-looking main result of the present paper:
\be
\partial_k\Gamma_k=-\frac{1}{2}\hat\partial_k{\textrm {Tr}}\left[\log\Gamma^{(2)}_{\bar\psi\psi}+\log\Gamma^{(2)}_{\psi^T\bar\psi^T}+\log\left(1-\Gamma^{(2)}_{\psi^T\psi}\Gamma^{(2)-1}_{\bar\psi\psi}\Gamma^{(2)}_{\bar\psi\bar\psi^T}\Gamma^{(2)-1}_{\psi^T\bar\psi^T}\right)\right].
\label{wetterich-factorized-explicit}
\ee
The first two terms on the right-hand side (the sum of the tracelogs of the two off-diagonal elements of $\Gamma^{(2)}$) would be the result of a "naive" computation.
  
The local potential approximation corresponds to working with the following generic Ansatz for the Euclidean effective action:
\be
\Gamma=\int d^dx\left\{\bar\psi(x)\sls\partial_E\psi(x)+U(I_p[\psi,\bar\psi])\right\}.
\ee
Here the spinorial and internal indices of the fermion fields are suppressed and the potential $U$ depends exclusively on a (sub)set $I_p$ of independent invariants formed with them.
The kinetic part is obtained from the Minkowskian Dirac action $iS_M^D\rightarrow -S_E^D$ by the formal transformations (applied for instance, also when dealing with fermions on a lattice) $ix^0=x^d_E, x^l=x^l_E,\gamma^0=\gamma_E^d, i\gamma^l=\gamma^l_E,~l=1,...,d-1$, resulting in the anticommutation relations $\{\gamma^m_E,\gamma^n_E\}=2\delta_{mn},~~ m,n=1,...,d$. The Fourier transform of the Euclidean fields is defined by
\be
\psi(x_E)=\int \frac{d^dq_E}{(2\pi)^d}e^{iq_Ex_E}\psi(q_E).
\ee
Below we omit the lower index $E$ from all quantities.

In the next sections we compute the functional matrix $\Gamma^{(2)}$ from a LPA Ansatz for the Gross-Neveu and the Nambu--Jona-Lasinio model and derive the explicit expressions for the right-hand side of the Wetterich equation in the form (\ref{wetterich-factorized-explicit}) for both models.

\section{LPA FOR THE GROSS-NEVEU MODEL}

The model was introduced for $N_f$ flavors of fermionic species by the action \cite{gross74}
\be
S[\bar\psi,\psi]=\int d^4x\left\{\sum_{j=1}^{N_f}\bar\psi_j\sls\partial\psi_j+\frac{g}{2N_f}\left(\sum_{j=1}^{N_f}\bar\psi_j\psi_j\right)^2\right\}.
\ee
The LPA of the model is formulated with the Ansatz:
\be
\Gamma[\bar\psi,\psi]=\int d^4x\left\{\sum_{j=1}^{N_f}\bar\psi_j\sls\partial\psi_j+U\left[\left(\sum_{j=1}^{N_f}\bar\psi_j\psi_j\right)^2\right]\right\}.
\ee
There are several quartic invariants in this model built from a wider set of fermionic bilinears which transform nontrivially under Lorentz and internal transformations. We shall evaluate the functional trace on a general fermionic background. It will be shown that only flavor-independent bilinears are generated along the RG flow. Even then various Lorentz structures might appear and eventually also one has to take into account relations among them resulting from Fierz's identities \cite{jaeckel03}. This means that possibly more independent invariants will be generated in the RG flow, than the single one displayed in the argument of the effective potential.

Below we omit the explicit indication of the flavor sum when writing the matrix elements of the second derivative $\Gamma^{(2)}$ in the fermionic background $(\bar\psi_i,\psi_i)$:
\bea
&
\displaystyle
\Gamma^{(2)}_{\bar\psi_i\psi_l}=\left[\sls\partial\delta(x-y)+2U'(\bar\psi\psi)\right]\delta_{il}+\tilde U\psi_i\bar\psi_l\equiv G^{(0)-1}(x-y)\delta_{il}+\tilde U\psi_i\bar\psi_l,\nonumber\\
&\displaystyle
\Gamma^{(2)}_{\psi_i^T\bar\psi_l^T}=\left[\sls\partial\delta(x-y)-2U'(\bar\psi\psi)\right]\delta_{il}+\tilde U\bar\psi_i^T\psi_l^T\equiv G^{(T0)-1}(x-y)\delta_{il}+\tilde U\bar\psi_i^T\psi_l^T, \nonumber\\
&\displaystyle
\Gamma^{(2)}_{\bar\psi_i\bar\psi_l^T}=-\tilde U\psi_i\psi_l^T\qquad \Gamma^{(2)}_{\psi_i^T\psi_l}=-\tilde U\bar\psi_i^T\bar\psi_l
\label{Gamma-2-GN}
\eea
with
\be
(\bar\psi\psi)\equiv\sum_{j=1}^{N_f}\bar\psi_j\psi_j,\qquad I_{GN}\equiv (\bar\psi\psi)^2,\qquad \tilde U=2U'+4I_{GN}U''.
\ee
In the brackets summation is also understood over the spinor indices; therefore, the combination $(\bar\psi\psi)$ is a constant of bosonic nature. It corresponds to the auxiliary field usually introduced by the Hubbard-Stratonovich transformation. Following \cite{gies02} we assume that the compositeness scale of this condensate is much higher than the scale $k$ relevant to the actual physics, therefore, this combination will be treated as an elementary scalar.

 The matrix elements of the inverse matrices can be given as a sum  over the respective geometric series in the bilinear form of the zeroth order propagators, $G^{(0)}$ and $G^{(T0)}$ :
\bea
  && \left(\Gamma^{(2)}_{\bar\psi\psi}\right)^{-1}_{i\ell} =  G^{(0)}
  \delta_{i\ell} - \frac{\tilde U G^{(0)} \psi_i\bar\psi_\ell
    G^{(0)}}{1 + \tilde U \bar\psi_k G^{(0)}\psi_k},\nonumber\\
  && \left(\Gamma^{(2)}_{\psi^T\bar\psi^T}\right)^{-1}_{i\ell} =  G^{(T0)}
  \delta_{i\ell} - \frac{\tilde U G^{(T0)} \bar\psi_i^T\psi_\ell^T
    G^{(T0)}}{1 + \tilde U \psi_k^T G^{(T0)}\bar\psi_k^T}.
\label{inverse-in-series}
\eea
These formulas are derived in some detail in Appendix B [see Eq.(\ref{propagator-inversion})]. 
The relevant third term on the right-hand side of (\ref{wetterich-factorized-explicit}) is found after evaluating the spinorial trace of each term in the Taylor series of the logarithm. Making use of the diagonal elements of $\Gamma^{(2)}$ appearing in the last line of (\ref{Gamma-2-GN}) one finds
\be
{\textrm{Tr}}\log\left(1-\Gamma^{(2)}_{\psi^T\psi}\Gamma^{(2)-1}_{\bar\psi\psi}\Gamma^{(2)}_{\bar\psi\bar\psi^T}\Gamma^{(2)-1}_{\psi^T\bar\psi^T}\right)=-\int_q\log\left(1-\tilde U^2(\bar\psi_l\Gamma^{(2)-1}_{\bar\psi_l\psi_k}\psi_k)(\psi^T_m\Gamma^{(2)-1}_{\psi^T_m\bar\psi^T_i}\bar\psi_i^T)\right).
\label{correction-tracelog}
\ee
Then one finds for the matrix elements in (\ref{correction-tracelog}) the following expressions:
\be
\bar\psi_\ell \left(\Gamma^{(2)}_{\bar\psi\psi}\right)^{-1}_{\ell
    k} \psi_k = \bar\psi_k G^{(0)}\psi_k - \frac{\tilde U
    \bar\psi_\ell G^{(0)} \psi_\ell \bar\psi_k G^{(0)}\psi_k}{1 +
    \tilde U \bar\psi_k G^{(0)}\psi_k} = \frac{\bar\psi_k
    G^{(0)}\psi_k}{1 + \tilde U \bar\psi_k G^{(0)}\psi_k}
\label{GN-propagator-1}
\ee
and
\be
\psi_m^T \left(\Gamma^{(2)}_{\psi^T\bar\psi^T}\right)^{-1}_{mi}
  \bar\psi_i^T = \psi_m^T G^{(T0)}\bar\psi_i^T - \frac{\tilde U
    \psi_m^T G^{(T0)} \bar\psi_m^T\psi_i^T G^{(T0)}\bar\psi_i^T}{1 +
    \tilde U \psi_k^T G^{(T0)}\bar\psi_k^T}  = \frac{\psi_i^T
    G^{(T0)}\bar\psi_i^T}{1 + \tilde U \psi_k^T
    G^{(T0)}\bar\psi_k^T}.
\label{GN-propagator-2}
\ee
One recognizes that there is no flavor mixing in the matrix elements; therefore, in the final result only flavor invariant bilinears of the spinors will appear. One can rewrite the above quantities in a simpler form where the $N_f$ factors, important from the point of view of correct $N_f$ scaling, are explicitely displayed: 
\be
\bar\psi_\ell \left(\Gamma^{(2)}_{\bar\psi\psi}\right)^{-1}_{\ell
    k} \psi_k =\frac{N_f(\bar\zeta G^{(0)}\zeta)}{1+\tilde U N_f(\bar\zeta G^{(0)}\zeta)},
\qquad
\bar\psi_\ell \left(\Gamma^{(2)}_{\bar\psi\psi}\right)^{-1}_{\ell
    k} \psi_k=\frac{N_f(\zeta^T G^{(T0)}\bar\zeta^T)}{1+\tilde U N_f(\zeta^T G^{(T0)}\bar\zeta^T)},
\ee
where $\zeta$ is a normalized spinor representing any of the flavors. With this, one can express the invariant as $I_{GN}=N_f^2(\bar\zeta\zeta)^2$.
The bilinears of the zeroth order propagators have the following expressions ($m_\psi=2U'N_f(\bar\zeta\zeta)$):
\be
\bar\zeta G^{(0)}\zeta=\frac{-i\bar\zeta\sls q\zeta+m_\psi\bar\zeta\zeta}{q^2+m_\psi^2},\qquad \zeta^T G^{(T0)}\bar\zeta^T=\frac{i\bar\zeta\sls q\zeta+m_\psi\bar\zeta\zeta}{q^2+m_\psi^2}.
\label{zeroth-bilinears}
\ee

The expressions for the various pieces on the right-hand side of (\ref{wetterich-factorized-explicit}) for the GN model are given, respectively, in terms of the quantities evaluated above as follows:
\bea
{\textrm {Tr}}\left(\log \tilde \Gamma^{(2)}_{\bar\psi_i\psi_p}+\log \tilde \Gamma^{(2)}_{\psi_i^T\bar\psi_p^T}\right)&=& {\textrm {Tr}}\left(\log  (G^{(0)-1}\delta_{ip})+\log  (G^{(T0)-1}\delta_{ip})\right)\nonumber\\
&&-\int_q\left(\log[1+\tilde U N_f(\bar\zeta G^{(0)}\zeta)]+\log[1+\tilde U N_f(\zeta^T G^{(T0)}\bar\zeta^T)]\right),\nonumber\\
{\textrm{Tr}}\log\left(\delta_{in}-\Gamma^{(2)}_{\psi^T_i\psi_k} \Gamma^{(2)-1}_{\bar\psi_k\psi_l}\Gamma^{(2)}_{\bar\psi_l\bar\psi^T_m} \Gamma^{(2)-1}_{\psi^T_m\bar\psi^T_n}\right)&=&
-\int_q\log\left(1-\frac{\tilde U^2N_f^2(\bar\zeta G^{(0)}\zeta)(\zeta^T\tilde G^{(T0)}\zeta^T)}{(1+\tilde UN_f(\bar\zeta G^{(0)}\zeta))(1+(\zeta^T\tilde G^{(T0)}\zeta^T))}\right).
\label{tracelog-pieces}
\eea
Here $\textrm{Tr}$ refers to the flavor and spinor (Dirac) indices plus integration over the contribution from the different Fourier modes. On the right-hand sides there are terms where the discrete traces were already formed and only the $q$ integration is left behind.

Using (\ref{tracelog-pieces}) with the explicit expressions of the zeroth order propagators (\ref{zeroth-bilinears}), one finds, after some elementary algebra
\bea
{\textrm {Tr}}\left(\log \tilde \Gamma^{(2)}_{\bar\psi_i\psi_p}+\log \tilde \Gamma^{(2)}_{\psi_i^T\bar\psi_p^T}\right)&
=& N_f{\textrm {Tr}_{D,q}}\left(\log G^{(0)-1}+\log  G^{(T0)-1}\right)\nonumber\\
&&-\int_q\log\frac{\left[(q^2+m_\psi^2+\tilde U N_fm_\psi \bar\zeta\zeta)^2+\tilde U^2N_f^2(\bar\zeta\sls q\zeta)^2\right]}{(q^2+m_\psi^2)^2},
\label{tracelog-1}
\eea
where $\textrm {Tr}_{D,q}$ refers only to the trace over the Dirac-matrix structure and the $q$ integration, and
\bea
&{\textrm{Tr}}\log&\left(\delta_{in}-\Gamma^{(2)}_{\psi^T_i\psi_k} \Gamma^{(2)-1}_{\bar\psi_k\psi_l}\Gamma^{(2)}_{\bar\psi_l\bar\psi^T_m} \Gamma^{(2)-1}_{\psi^T_m\bar\psi^T_n}\right)\nonumber\\
&&=-\int_q\log\left(1-\frac{N_f^2\tilde U^2\left[(\bar\zeta\sls q\zeta)^2+m_\psi^2(\bar\zeta\zeta)^2\right]}{\left[(q^2+m_\psi^2+\tilde U N_fm_\psi \bar\zeta\zeta)^2+\tilde U^2N_f^2(\bar\zeta\sls q\zeta)^2\right]}\right)\nonumber\\
&&=-\int_q\log\frac{(q^2+m_\psi^2+\tilde U N_fm_\psi \bar\zeta\zeta)^2-N_f^2\tilde U^2m_\psi^2(\bar\zeta\zeta)^2}{\left[(q^2+m_\psi^2+\tilde U N_fm_\psi \bar\zeta\zeta)^2+\tilde U^2N_f^2(\bar\zeta\sls q\zeta)^2\right]}.
\label{tracelog-2}
\eea
It is very remarkable that in the sum of the two contributions the terms containing $\bar\zeta\sls q\zeta$ cancel. In this way we automatically obtain, on the right hand side of the RGE, an expression that depends only on $I_{GN}$:
\be
 \partial_k\Gamma_k=-\frac{1}{2}\int_q\hat\partial_k\left[(4N_f+1)\log(q^2+4U'^2I_{GN})-\log(q^2+4U'(U'+\tilde U)I_{GN})\right].
\label{GN-Wetterich-1}
\ee

Next one replaces $q^2$ everywhere by $q_R^2=q^2(1+r_\psi)^2$, where the "linear" suppression function for the modes below the scale $k$ is used \cite{litim64}:
\be
r_\psi=\left(\frac{k}{\sqrt{q^2}}-1\right)\Theta(k^2-q^2).
\label{litim-suppression}
\ee 
With this choice one can trivially perform  the integration over $q$ on the right-hand side of (\ref{GN-Wetterich-1})  after the $\hat\partial_k$ derivative is
calculated:
\be
\partial_k\Gamma_k=-\frac{k^{d+1}S_d}{d(2\pi)^d}\left[(4N_f+1)\frac{1}{k^2+4U'^2I_{GN}}
-\frac{1}{k^2+4U'(U'+\tilde U)I_{GN}}\right],
\label{GN-FRGE}
\ee
where $S_d$ is the surface of the $d$-dimensional unit sphere. The structure of this equation is rather similar to that obtained in the partially bosonized formulation for the potential depending on the scalar auxiliary field: the leading large-$N_f$ contribution from the fermions is corrected by a term of opposite sign which can be interpreted as the contribution of a bosonic propagating mode.

The RG flow in the leading large-$N_f$ approximation is determined by the first term on the right-hand side. One expands both sides in power series with respect to $I_{GN}$ and comparing the coefficients of the linear terms one finds (the constant piece presents no interest)
\be
\partial_k U'=\frac{16N_fk^{d-3}}{d(2\pi)^d}S_dU'^2.
\ee
Choosing the usual notation $U'=\bar g/2N_f$ (see, for instance Ref.\cite{braun11}) and introducing the dimensionless coupling $g=k^{d-2}\bar g$ one arrives at the well-known perturbative large-$N_f$ flow:
\be
k\partial_kg=(d-2)g+\frac{4S_d}{d(2\pi)^d}g^2.
\ee
The right-hand side has a zero for some negative value of $g$. This is the same fixed point as discussed in \cite{braun11} but one has to take into account that the authors of \cite{braun11} use a different Euclidean spinor convention \cite{wetterich11} and that our adjoint spinor $\bar \psi$ has an extra factor $i$ relative to what they are using. As a consequence our coupling $g$ has an extra minus sign relative to the coupling used in the fermionic RG calculations following the convention of \cite{wetterich11}.

This fixed point which is a nontrivial example of the asymptotic safety of a perturbatively nonrenormalizable field theory can be investigated with the present formulation beyond the large-$N_f$ approximation and without restricting the discussion to the lowest powers in the expansion of $U(I_{GN})$. One has to look for the (non-Gaussian) stationary solutions of (\ref{GN-FRGE}) and perform the Wilsonian analysis around this fixed point.

\section{LPA FOR THE $N_f=1$ NAMBU--JONA-LASINIO MODEL}

The LPA Ansatz for the effective action for the one-flavor NJL model \cite{nambu61} is of the following form:
\be
\Gamma=\int d^4x[\bar\psi\sls\partial\psi+U(I_{NJL})],
\label{NJL-action}
\ee
where the $U_L(1)\times U_R(1)$ invariant variable of the potential is defined as
\be
I_{NJL}=(\bar\psi_R\psi_L)(\bar\psi_L\psi_R)=\frac{1}{4}\left((\bar\psi\psi)^2-(\bar\psi\gamma_5\psi)^2\right).
\ee
The calculation below will be done with a Grassmannian background $\zeta,\bar\zeta$ for which we impose $\bar\zeta\gamma_5\zeta=0$ simplifying the identification of the $I_{NJL}$ dependence on the right-hand side of the RG equation (\ref{wetterich-factorized-explicit}). We do not consider the dependence of the effective potential on other quartic invariants. The explicit expressions for the elements of the second functional derivative matrix of (\ref{NJL-action}) read as
\bea
\Gamma^{(2)}_{\psi^T\psi}=-\tilde U\bar\zeta^T\bar\zeta+\frac{1}{2}U'\bar\zeta_5^T\bar\zeta_5,&&
\Gamma^{(2)}_{\bar\psi\bar\psi^T}=-\tilde U\zeta\zeta^T+\frac{1}{2}U'\zeta_5\zeta_5^T,\nonumber\\
\Gamma_{\psi^T\bar\psi^T}^{(2)}=i\sls q^T-m_\psi+\tilde U\bar\zeta^T\zeta^T-\frac{1}{2}U'\bar\zeta_5^T\zeta_5^T&\equiv&
G_\psi^{(T0)-1}+\tilde U\bar\zeta^T\zeta^T-\frac{1}{2}U'\bar\zeta_5^T\zeta_5^T,\nonumber\\
\Gamma_{\bar\psi\psi}^{(2)}=i\sls q+m_\psi+\tilde U\zeta\bar\zeta-\frac{1}{2}U'\zeta_5\bar\zeta_5&
\equiv& G_\psi^{(0)-1}+\tilde U\zeta\bar\zeta-\frac{1}{2}U'\zeta_5\bar\zeta_5.
\label{NJL-Gamma2}
\eea
The following shorthand notations were introduced into the above expressions:
\bea
&\displaystyle
\zeta_5=\gamma_5\zeta, \quad \bar\zeta_5=\bar\zeta\gamma_5,\quad m_\psi=\frac{1}{2}U'(\bar\zeta\zeta),\quad U'=\frac{dU}{dI_{NJL}},\nonumber\\
&\displaystyle
\tilde U=\frac{1}{2}U'+U''I_{NJL}.
\eea

We focus first on the evaluation of 
${\textrm{Tr}_{D,q}}\log\left(I-\Gamma^{(2)}_{\psi^T\psi}\Gamma^{(2)-1}_{\bar\psi\psi}\Gamma^{(2)}_{\bar\psi\bar\psi^T}\Gamma^{(2)-1}_{\psi^T\bar\psi^T}\right)$ on the right-hand side of (\ref{wetterich-factorized-explicit}). The $n$th term of the power expansion of the logarithm, 
\be
{\textrm{Tr}_{D,q}}\left[\left(-\tilde
    U\bar\zeta^T\bar\zeta+\frac{1}{2}U'\bar\zeta_5^T\bar\zeta_5\right)\Gamma^{(2)-1}_{\bar\psi\psi}\left(-\tilde
    U\zeta\zeta^T+\frac{1}{2}U'\zeta_5\zeta_5^T\right)\Gamma^{(2)-1}_{\psi^T\bar\psi^T}\right]^n
\label{eq:NJLtrace}
\ee
can be interpreted as $n$ repetitions of two subsequent "transfers." The first step consists of the transfer of the Grassmann variables $\zeta,\zeta_5$ with $\Gamma^{(2)-1}_{\bar\psi\psi}$ to $\bar\zeta,\bar\zeta_5$, followed by the second transfer, now of $\bar\zeta^T, \bar\zeta^T_5$ with $\Gamma^{(2)-1}_{\psi^T\bar\psi^T}$ to $\zeta^T,\zeta^T_5$. The transfer in the $n$th step closes back on the starting "configuration" since one computes the Dirac trace. This last step also results in a negative sign, since the leftmost Grassmann spinor has to be moved to the end of the expression. The formula expressing this is
\bea
&&{\textrm{Tr}_{D,q}}\left[\left(-\tilde U\bar\zeta^T\bar\zeta+\frac{1}{2}U'\bar\zeta_5^T\bar\zeta_5\right)\Gamma^{(2)-1}_{\bar\psi\psi}\left(-\tilde U\zeta\zeta^T+\frac{1}{2}U'\zeta_5\zeta_5^T\right)\Gamma^{(2)-1}_{\psi^T\bar\psi^T}\right]^n\nonumber\\
&&
=-{\textrm{Tr}_{2,q}}\left[
\begin{pmatrix}
{\kappa \Gamma^{-1}_{00}} & {\sqrt{\kappa\delta}}\Gamma^{-1}_{05}\\
{\sqrt{\kappa\delta}}\Gamma^{-1}_{50}&
{\delta \Gamma^{-1}_{55}}
\end{pmatrix}
\cdot
\begin{pmatrix}
{\kappa \Gamma_{00}^{(T)-1}} & {\sqrt{\kappa\delta}}\Gamma_{05}^{(T)-1}\\
{\sqrt{\kappa\delta}}\Gamma_{50}^{(T)-1}&
{\delta \Gamma_{55}^{(T)-1}}
\end{pmatrix}\right]^n.
\label{transfer-rewriting}
\eea
In this equality the Dirac trace is replaced by tracing the product of $2\times 2$ matrices.
In the above expression we introduce the following abbreviations:
\bea
&
\Gamma^{-1}_{00}=\bar\zeta\Gamma^{(2)-1}_{\bar\psi\psi}\zeta,\quad \Gamma^{-1}_{05}=\bar\zeta\Gamma^{(2)-1}_{\bar\psi\psi}\zeta_5,\quad \Gamma^{-1}_{50}=\bar\zeta_5\Gamma^{(2)-1}_{\bar\psi\psi}\zeta,\quad \Gamma^{-1}_{55}=\bar\zeta_5\Gamma^{(2)-1}_{\bar\psi\psi}\zeta_5,\nonumber\\
&
\Gamma_{00}^{(T)-1}=\zeta^T\Gamma^{(2)-1}_{\psi^T\bar\psi^T}\bar\zeta^T,\quad \Gamma_{05}^{(T)-1}=\zeta^T\Gamma^{(2)-1}_{\psi^T\bar\psi^T}\bar\zeta_5^T,\quad \Gamma_{50}^{(T)-1}=\zeta_5^T\Gamma^{(2)-1}_{\psi^T\bar\psi^T}\bar\zeta^T,\quad \Gamma_{55}^{(T)-1}=\zeta_5^T\Gamma^{(2)-1}_{\psi^T\bar\psi^T}\bar\zeta_5^T,\nonumber\\
&
\kappa=\tilde U,\qquad \delta=-\frac{1}{2}U'.
\eea
Using this result one rewrites the tracelog into a "logdet" form: 
\bea
&&{\textrm{Tr}}\log\left(I-\Gamma^{(2)}_{\psi^T\psi}\Gamma^{(2)-1}_{\bar\psi\psi}\Gamma^{(2)}_{\bar\psi\bar\psi^T}\Gamma^{(2)-1}_{\psi^T\bar\psi^T}\right)\nonumber\\
&=&-\int_q\log\det
\left\{I-\left[
\begin{pmatrix}
{\kappa \Gamma^{-1}_{00}} & {\sqrt{\kappa\delta}}\Gamma^{-1}_{05}\\
{\sqrt{\kappa\delta}}\Gamma^{-1}_{50}&
{\delta \Gamma^{-1}_{55}}
\end{pmatrix}
\cdot
\begin{pmatrix}
{\kappa \Gamma_{00}^{(T)-1}} & {\sqrt{\kappa\delta}}\Gamma_{05}^{(T)-1}\\
{\sqrt{\kappa\delta}}\Gamma_{50}^{(T)-1}&
{\delta \Gamma_{55}^{(T)-1}}
\end{pmatrix}\right]\right\}
\label{transfer-matrix-rewriting-1}
\eea
[more details on the derivation can be found in Appendix B, see (\ref{leads-to-41})]. 
Following the same line of operations one finds, for the first two terms in the sum on the right-hand side of (\ref{wetterich-factorized-explicit})
\bea
{\textrm{Tr}}\log\Gamma^{(2)}_{\bar\psi\psi}&=&{\textrm{Tr}}\log G^{(0)-1}-\int_q\log\det(I+{\cal A}),\nonumber\\
{\textrm{Tr}}\log\Gamma^{(2)}_{\psi^T\bar\psi^T}&=&{\textrm{Tr}}\log G^{(T0)-1}-\int_q\log\det(I+{\cal A^{(T)}}),
\label{naive-propagators}
\eea
where the following matrices defined with the matrix elements of the massive free fermion propagators will appear:
\be
{\cal A}= \begin{pmatrix}
{\kappa G_{00}^{(0)}} & {\sqrt{\kappa\delta}}G_{05}^{(0)}\\
{\sqrt{\kappa\delta}}G_{50}^{(0)}&
{\delta G_{55}^{(0)}}
\end{pmatrix},\qquad
{\cal A^{(T)}}=\begin{pmatrix}
{\kappa G_{00}^{(T0)}} & {\sqrt{\kappa\delta}}G_{05}^{(T0)}\\
{\sqrt{\kappa\delta}}G_{50}^{(T0)}&
{\delta G_{55}^{(T0)}}
\end{pmatrix}.
\ee
With their help one also finds the matrix elements of
$\Gamma^{(2)-1}_{\bar\psi\psi}, \Gamma^{(2)-1}_{\psi^T\bar\psi^T}$
needed for the evaluation of (\ref{transfer-matrix-rewriting-1}) [see
Eq.(\ref{geometric-sum}) in Appendix B]:
\be
\label{eq:NJLinverse}
u_{ij}\Gamma_{ij}^{-1}=\delta_{ij}-\left[(I+{\cal A})^{-1}\right]_{ij},
\qquad
u_{ij}^{(T)}\Gamma_{ij}^{(T)-1}=\delta_{ij}-\left[(I+{\cal A^{(T)}})^{-1}\right]_{ij},\qquad i,j=0,5
\ee
(no summation is understood on the repeating indices) and
\be
u_{00}=u_{00}^{(T)}=\kappa,\quad u_{05}=u_{50}=u_{05}^{(T)}=u_{50}^{(T)}=\sqrt{\kappa\delta}, \quad u_{55}=u_{55}^{(T)}=\delta.
\ee
These expressions can be used to rewrite the argument of the logdet in (\ref{transfer-matrix-rewriting-1}) using the shorthand notations $B^{-1}=(I+{\cal A})^{-1},~B^{(T)-1}=(I+{\cal A^{(T)}})^{-1}$:
\be
I-(I-B^{-1})(I-B^{(T)-1})=B^{-1}(I+{\cal A}+{\cal A^{(T)}})B^{(T)-1},
\ee
which gives
\be
{\textrm{Tr}}\log\left(1-\Gamma^{(2)}_{\psi^T\psi}\Gamma^{(2)-1}_{\bar\psi\psi}\Gamma^{(2)}_{\bar\psi\bar\psi^T}\Gamma^{(2)-1}_{\psi^T\bar\psi^T}\right)=\int_q\left[\log\det(I+{\cal A})+\log\det(I+{\cal A^{(T)}})-\log\det(I+{\cal A}+{\cal A^{(T)}})\right].
\ee
Adding this to the contributions from the other two terms appearing in (\ref{naive-propagators}), one observes considerable cancellation again:
\be
\partial_k\Gamma_k=-\frac{1}{2}\hat\partial_k\left[{\textrm{Tr}}\log G^{(0)-1}+{\textrm{Tr}}\log G^{(T0)-1}-\int_q\log\det(I+{\cal A}+{\cal A^{(T)}})\right].
\ee

Finally, one can reexpress the right side of this expression with the help of the matrix elements of the zeroth order propagators:
\bea
&\displaystyle
(q^2+m_\psi^2)G_{00}^{(0)}=m_\psi\bar\zeta\zeta-i\bar\zeta\sls q\zeta,\quad 
(q^2+m_\psi^2)G_{55}^{(0)}=m_\psi\bar\zeta\zeta+i\bar\zeta\sls q\zeta,\nonumber\\
&\displaystyle
(q^2+m_\psi^2)G_{05}^{(0)}=-(q^2+m_\psi^2)G_{50}^{(0)}=-i\bar\zeta\sls q\gamma_5\zeta\nonumber\\
&\displaystyle
(q^2+m_\psi^2)G_{00}^{(T0)}=m_\psi\bar\zeta\zeta+i\bar\zeta\sls q\zeta,\quad
(q^2+m_\psi^2)G_{55}^{(T0)}=m_\psi\bar\zeta\zeta-i\bar\zeta\sls q\zeta\nonumber\\
&\displaystyle
(q^2+m_\psi^2)G_{05}^{(T0)}=-(q^2+m_\psi^2)G_{50}^{T0}=i\bar\zeta\sls q\gamma_5\zeta.
\label{zeroth-propagator-matrix}
\eea
One finds [using the abbreviated notation $q_R^2=q^2(1+r_\psi)^2$ for the momentum in the $k$-suppressed propagators]
\be
\partial_k\Gamma_k=-\frac{1}{2}\hat\partial_k\int_q\left[6\log(q_R^2+m_\psi^2)-\log\left(q^2_R+m_\psi^2+2\kappa m_\psi\bar\zeta\zeta\right)
\left(q^2_R+m^2_\psi+2\delta m_\psi\bar\zeta\zeta\right)\right].
\ee
In the final step one makes use of the linear suppression function (\ref{litim-suppression}) and performs the $q$ integration. The right-hand side of the resulting equation is expressed through the invariant $I_{NJL}$:
\be
\partial_kU_k=-\frac{k^{d+1}S_d}{(2\pi)^d}\left[\frac{6}{k^2+U'^2I_{NJL}}-\frac{1}{k^2+(3U'^2+4U'U''I_{NJL})I_{NJL}}-\frac{1}{k^2-U'^2I_{NJL}}\right].
\ee

\section{CONCLUSIONS}

In this paper we achieved considerable technical progress in including higher dimensional fermionic operators in the effective action and in deriving the RGE in LPA for the corresponding effective potential. Somewhat unexpectedly there was no need to take into account Fierz's identities, either in the Gross-Neveu model or in the one-flavor Nambu--Jona-Lasinio model. No flavor-dependent fermionic bilinears were generated in the RG flow of the Gross-Neveu model. This should be contrasted with the experience gained with Taylor-expanded functional determinants \cite{gies02}. 
 
The resulting FRG equations are of comparable simplicity to those of the bosonic models like the linear sigma model. In both models investigated by us the contribution from the massive fermion loop is corrected by terms that can be interpreted as a sort of bosonic loop contribution. This structure is rather similar to the structure of the RGE arising for the effective potential of the auxiliary bosonic variable in partially bosonized reformulations. There is definite hope that the proposed procedure can also handle  quark-meson models of phenomenological interest\cite{jakovac13}. This will give us the chance to test the stability of the conclusions drawn with FRG investigations based on simpler Ans\"atze \cite{schaefer05,schaefer07,braun10,herbst11} concerning the nature of the finite temperature phase transition in this class of effective models of QCD. 
The fundamental question is if the factorized form of the functional determinant of the field fluctuations constructed in the present paper for the pure fermionic theories can also be achieved in the case when the interaction of the fermi fields with the bosons is switched on. 

We emphasize that the treatment of the fermionic action is purely algebraic, so the result is independent of the analytic properties of the model such as dimensionality of the spacetime, momentum dependence of the fermionic kernels, or spacetime signature. In particular, one can also have  Minkowskian spacetime signature. Then one can either use the Euclidean result with analytic continuation by applying an appropriate kernel for the suppression functional, which
obeys the general analyticity requirements \cite{floerchinger12, Kamikado:2013sia}, or one can work directly in Minkowski space, where the formulation of the FRG equations is based on the 2-particle-irreducible methods \cite{Berges:2012ty, Jakovac:2006gi}.

Also, one should investigate Ans\"atze which imply non-trivial wave function renormalisation (the next term of the gradient expansion). In some cases where partial bosonization is applied, the wave function renormalization of the auxiliary fields (dynamical scaling of their kinetic Lagrangian) has an essential impact on the fixed point structure of the theory \cite{scherer12a}.

\section*{Acknowledgements}
This research was supported by the Hungarian Research Fund under Contracts No. K-104292 and No. K-77534.

\appendix

\renewcommand{\theequation}{A.\arabic{equation}}
\setcounter{equation}{0}  

\section*{Appendix A: ANALYSIS OF THE CONSISTENCY CONDITION}
\label{sec:appa}

In Eq.~\eqref{consistency} we determined the consistency equations for
the factorization of the fermion propagator in a Grassmannian
background field, cf. Eq.~\eqref{factorisation}. In this appendix we
present the solution of this equation, assuming that all the
$\Gamma^{(2)}$ symbols represent invertible matrices.\footnote{If this was not the case, we could introduce a regularization parameter that makes these matrices invertible. In the end result, we may put
  the regularization parameter to zero.}

First multiply the first equation of \eqref{consistency} by $\tilde
G_\psi^{(T)}$ from the left, and by $\Gamma^{(2)-1}_{\bar\psi\bar\psi^T}$
from the right. Similarly, multiply the second equation by $G_\psi$ from the
right, and by $\Gamma^{(2)-1}_{\bar\psi\bar\psi^T}$ from the left. We find
\begin{eqnarray}
  \label{eq:A1}
  {\tilde G_\psi}^{(T)} \Gamma^{(2)}_{\psi^T\bar\psi^T}
  \Gamma^{(2)-1}_{\bar\psi\bar\psi^T}  &=&
  \Gamma^{(2)-1}_{\bar\psi\bar\psi^T} + \frac14 {\tilde 
    G_\psi}^{(T)} \Gamma^{(2)}_{\psi^T\psi} {\tilde G_\psi}\nonumber\\ 
  \Gamma^{(2)-1}_{\bar\psi\bar\psi^T} \Gamma^{(2)}_{\bar\psi\psi} {\tilde
    G_\psi}&=& \Gamma^{(2)-1}_{\bar\psi\bar\psi^T} + \frac14 {\tilde
    G_\psi}^{(T)} \Gamma^{(2)}_{\psi^T\psi} {\tilde G_\psi}.
\end{eqnarray}
Since the right-hand sides of both equations are the same, we find
${\tilde G_\psi}^{(T)} \Gamma^{(2)}_{\psi^T\bar\psi^T}
\Gamma^{(2)-1}_{\bar\psi\bar\psi^T} = \Gamma^{(2)-1}_{\bar\psi\bar\psi^T}
\Gamma^{(2)}_{\bar\psi\psi} {\tilde G_\psi}$, or
\begin{equation}
  \label{eq:ref2}
  {\tilde G_\psi}^{(T)}  = \Gamma^{(2)-1}_{\bar\psi\bar\psi^T} \Gamma^{(2)}_{\bar\psi\psi}
  {\tilde G_\psi}\Gamma^{(2)}_{\bar\psi\bar\psi^T}\Gamma^{(2)-1}_{\psi^T\bar\psi^T}.
\end{equation}
After substituting this back into the first equation, we find
\begin{equation}
  \Gamma^{(2)}_{\bar\psi\psi} = {\tilde G_\psi}^{-1} + \frac14 \Gamma^{(2)}_{\bar\psi\psi}
  {\tilde G_\psi}\Gamma^{(2)}_{\bar\psi\bar\psi^T}\Gamma^{(2)-1}_{\psi^T\bar\psi^T}
  \Gamma^{(2)}_{\psi^T\psi}.
\end{equation}
We multiply this equation by $\Gamma^{(2)-1}_{\bar\psi\psi}$ from the
left, and introduce a new variable $U = {\tilde
  G_\psi}\Gamma^{(2)}_{\bar\psi\psi}$. Then we find
\begin{equation}
  1 = U^{-1} + \frac14 U
  \Gamma^{(2)-1}_{\bar\psi\psi}\Gamma^{(2)}_{\bar\psi\bar\psi^T}
  \Gamma^{(2)-1}_{\psi^T\bar\psi^T} \Gamma^{(2)}_{\psi^T\psi}.
\end{equation}
With the notation $M = \Gamma^{(2)-1}_{\bar\psi\psi}
\Gamma^{(2)}_{\bar\psi\bar\psi^T} \Gamma^{(2)-1}_{\psi^T\bar\psi^T}
\Gamma^{(2)}_{\psi^T\psi}$, this can be written as
\begin{equation}
  \label{eq:ref1}
  1 = U^{-1} + \frac14 U M \quad\Rightarrow\quad U = 1 + \frac14 U M U.
\end{equation}

Now we multiply the equation by $M^{1/2}$. both from the left and the right, to obtain
\begin{equation}
  M^{1/2} U M^{1/2} = M + \frac14 M^{1/2} U M U M^{1/2} = M + \frac14 (M^{1/2} U M^{1/2})^2.
\end{equation}
Completing the square this leads to
\begin{equation}
  1-M = \left(\frac12 M^{1/2} U M^{1/2} -1\right)^2,
\end{equation}
which has the solution
\begin{equation}
  U = 2\left(1\pm \sqrt{1-M}\right) M^{-1}.
\end{equation}
Therefore
\begin{equation}
  {\tilde G_\psi} = 2 \left(1\pm \sqrt{1-M}\right) M^{-1}
  \Gamma^{(2)-1}_{\bar\psi\psi}.
\end{equation}
We choose the negative sign, since we remark that for $M\to 0$ the solution
of \eqref{eq:ref1} is $U=1$ or ${\tilde
  G_\psi}=\Gamma^{(2)-1}_{\bar\psi\psi}$. Using the actual form of $M$ we find
\begin{equation}
  {\tilde G_\psi} = 2\left(1 - \sqrt{1 -
      \Gamma^{(2)-1}_{\bar\psi\psi}\Gamma^{(2)}_{\bar\psi\bar\psi^T}
      \Gamma^{(2)-1}_{\psi^T\bar\psi^T} \Gamma^{(2)}_{\psi^T\psi}}
  \right) \Gamma^{(2)-1}_{\psi^T\psi} \Gamma^{(2)}_{\psi^T\bar\psi^T}
  \Gamma^{(2)-1}_{\bar\psi\bar\psi^T}.
\end{equation}
>From \eqref{eq:ref2} we can write
\begin{equation}
  {\tilde G_\psi}^{(T)} = 2\Gamma^{(2)-1}_{\bar\psi\bar\psi^T} \Gamma^{(2)}_{\bar\psi\psi}
  \left(1 - \sqrt{1 -
      \Gamma^{(2)-1}_{\bar\psi\psi}\Gamma^{(2)}_{\bar\psi\bar\psi^T}
      \Gamma^{(2)-1}_{\psi^T\bar\psi^T} \Gamma^{(2)}_{\psi^T\psi}} \right)
  \Gamma^{(2)-1}_{\psi^T\psi}.
\end{equation}
Finally, we can use in these expressions the identity
\begin{equation}
  \sqrt{1+AB}\, B^{-1} = B^{-1} \sqrt{1+BA}
\end{equation}
to arrive at the Eq. \eqref{eq:contr_solve}. This identity can be
proved by power expanding both sides.

It is worth mentioning another matrix identity that we repeatedly use in the main text in
factorizing the tracelog expressions:
\begin{equation}
  {\textrm {Tr}} \log AB = {\textrm {Tr}} \log A + {\textrm {Tr}} \log B.
\end{equation}
This is a direct consequence of the ${\textrm {Tr}} \log A=\log\det A$
identity.

\renewcommand{\theequation}{B.\arabic{equation}}
\setcounter{equation}{0}  

\section*{Appendix B: FUNCTIONS OF TENSOR PRODUCTS}
Several formulaS of thIS paper rely on the evaluation of some function
of tensor products of (Grassmann) vectors. So let us study the general
formula
\begin{equation}
  f(I) = \sum\limits_{n=0}^\infty f_n I^n,
\end{equation}
where
\begin{equation}
  I= \sum\limits_{a=1}^r \eta_a \xi^T_a.
\end{equation}
Here $\eta_a \xi^T_a$, as in the rest of this paper, is the shorthand
notation of the tensor product $\eta_a\otimes\xi^T_a$, while the
expression $\xi^T_a\eta_a$ implies the scalar product.

The number of components $r$ corresponds to the rank of the vector
combinations appearing in the second functional derivative matrix
$\Gamma^{(2)}$.  In the GN model we have $r=1$, but both $\eta_1$ and
$\xi_1$ have $N_f$ flavor components. In the formulas below we
should then identify $(\eta_1)_i = \psi_i$ and $(\xi_1^T)_i = \bar\psi_i$, or
$(\eta_1)_i = \bar\psi^T_i$ and $(\xi_1^T)_i = \psi^T_i$. In the NJL
model, on the other hand, we have $r=2$ [cf., for example
\eqref{eq:NJLtrace}]. The identification is $(\eta_1,\eta_2)=(\zeta,
\zeta_5)$ and $(\xi_1^T,\xi_2^T)=(\bar\zeta, \bar\zeta_5)$, or
$(\eta_1,\eta_2)=(\bar\zeta^T, \bar\zeta^T_5)$ and
$(\xi_1^T,\xi_2^T)=(\zeta^T, \zeta^T_5)$.

The feature that will help us is that $I^n$ for $n\ge1$ is of the
form
\begin{equation}
  I^n = \sum\limits_{a,b=1}^r (M^{n-1})_{ab} \eta_a \xi^T_b,
  \qquad\mathrm{where}\quad M_{ab} =   \xi_a^T\eta_b,
\end{equation}
and $M^n$ is meant in the sense of matrix multiplication. For $n=1$ this is
evident because $M^0=1$. For $n=2$ we check that this is the case by explicit calculation,
\begin{equation}
  I^2 = \sum\limits_{a,b=1}^r (\eta_a
  \xi^T_a)(\eta_b \xi^T_b) = \sum\limits_{a,b=1}^r
  (\xi_a^T\eta_b) (\eta_a \xi^T_b).
\end{equation}
For the general case one can prove the result by induction, assuming that the statement is true for a given $n$:
\begin{equation}
  I^{n+1} = \sum\limits_{a,b,c=1}^r \left[ (M^{n-1})_{ab}
    \eta_a \xi^T_b\right] [\eta_c
  \xi^T_c] = \sum\limits_{a,c=1}^r  (\eta_a \xi^T_c)
  \sum\limits_{b=1}^r(M^{n-1})_{ab}(\xi_b^T\eta_c).
\end{equation}

Any matrix valued function is defined by a power
series. If $f(x)=\sum_n f_n x^n$ is the series expansion, then the function $f(I)$ can be written in the following convenient form:
\begin{equation}
  \label{eq:Appmaster}
  f(I) = f_0 + \sum\limits_{n=1}^\infty f_n \sum\limits_{a,b=1}^r
  (M^{n-1})_{ab} \eta_a \xi^T_b = f_0 + \sum\limits_{a,b=1}^r
  \left(M^{-1}\tilde f(M)\right)_{ab} \eta_a \xi^T_b,
\end{equation}
where $\tilde f(x) = f(x) - f(0)$.

Applications are as follows.
\begin{itemize}
\item Inverse propagator:
  \begin{equation}
    (G^{-1} + \sum\limits_{a=1}^r \eta_a \xi^T_a )^{-1} =
    \left[ G^{-1} (1 + \sum\limits_{a=1}^r G \eta_a
      \xi^T_a)\right]^{-1} = (1 + \sum\limits_{a=1}^r \eta'_a 
    \xi^T_a)^{-1} G,
  \end{equation}
  where $\eta'=G\eta$. Now $f=1/(1+x)$; thus $\tilde f =-x/(1+x)$, and
  we have
  \begin{equation}
    (G^{-1} + \sum\limits_{a=1}^r \eta_a \xi^T_a )^{-1} = G -
    \sum\limits_{a,b=1}^r (1+M)^{-1}_{ab} G\eta_a
    \xi^T_b G,\qquad\mathrm{where}\quad M_{ab} = \xi^T_aG\eta_b.
\label{propagator-inversion}
  \end{equation}
  Equation \eqref{inverse-in-series} is a special case of this formula.
\item From \eqref{eq:Appmaster} we immediately see that
  \begin{equation}
    \xi^T_a f(I) \eta_b = \left(M f(M) \right)_{ab}.
  \end{equation}
  In particular, as a matrix
  \begin{equation}
    \xi^T (1+I)^{-1} \eta = M (1+M)^{-1} = 1- (1+M)^{-1}.
    \label{geometric-sum}  
  \end{equation}
  This equation is relevant in deriving \eqref{eq:NJLinverse}.

\item As a third application we can observe that if $f_0=0$, then
  \begin{equation}
    \mathrm{Tr} f(I) = \pm \mathrm{Tr}_M f(M),
  \end{equation}
  where the $\pm$ signs are valid for bosonic or fermionic vectors,
  respectively. Here, on the right-hand side the trace of the matrix
  $M$ is understood, while on the left-hand side the trace is over the
  spinorial indices. In particular, in the fermionic one-loop integral
  we encounter a formula [cf. left-hand side of
  \eqref{transfer-matrix-rewriting-1} with \eqref{NJL-Gamma2}]
  \begin{eqnarray}
    && \mathrm{Tr}\log\left(1 - (\sum\limits_{a=1}^r c_a \eta_{1a}
      \xi^T_{1a}) \Gamma^{(2)-1}_{\bar\psi\psi} (\sum\limits_{b=1}^r c_b
      \eta_{2b} \xi^T_{2b}) \Gamma^{(2)-1}_{\psi^T\bar\psi^T}
    \right) =\nonumber\\ && =\mathrm{Tr}\log\left(1 -
      \sum\limits_{a,b=1}^r c_ac_b (\xi^T_{1a}
      \Gamma^{(2)-1}_{\bar\psi\psi} \eta_{2b}) \eta_{1a} \xi^T_{2b}
      \Gamma^{(2)-1}_{\psi^T\bar\psi^T} \right).
  \end{eqnarray}
  This can be written in the standard form if we introduce
  \begin{equation}
    \eta_a' = \sqrt{c_a} \eta_{1a},\qquad 
    {\xi'}^T_a = \sum\limits_{b=1}^r (\sqrt{c_a}\xi_{1a}^T
    \Gamma^{(2)-1}_{\bar\psi\psi} \sqrt{c_b}\eta_{2b})
    \sqrt{c_b}\xi_{2b}^T \Gamma^{(2)-1}_{\psi^T\bar\psi^T}.
  \end{equation}
  Then the equation reads
  \begin{equation}
    \mathrm{Tr}\log\left(1 - \sum\limits_{a=1}^r \eta'_a
      {\xi'}^T_a\right) = -\log\det\left(1 - M'\right),
\label{leads-to-41}  
  \end{equation}
  where $M'_{ab} = {\xi'}^T_a\eta'_b = (M_{\bar\psi\psi}
  M_{\psi^T\bar\psi^T})_{ab}$ and
  \begin{equation}
    (M_{\bar\psi\psi})_{ab} =\sqrt{c_a}\xi_{1a}^T
    \Gamma^{(2)-1}_{\bar\psi\psi} \sqrt{c_b}\eta_{2b},
    \quad\mathrm{and}\quad  (M_{\psi^T\bar\psi^T})_{ab} =
    \sqrt{c_a}\xi_{2a}^T \Gamma^{(2)-1}_{\psi^T\bar\psi^T}
    \sqrt{c_b}\eta_{1b}.
  \end{equation}
  This proves the right-hand side of \eqref{transfer-matrix-rewriting-1}.

\end{itemize}

\end{document}